\documentstyle[aps,prb]{revtex}

\begin{document}
\draft

\title{ DELOCALIZATION IN CONTINUOUS DISORDERED SYSTEMS}
\author{
M. Hilke\footnote {Current address, Dpt. of electrical engineering, 
Princeton University, NJ-08544 Princeton, USA.}
 and J. C. Flores\footnote{ Current address, Universidad de Chile FCFM
Dept. de F\'{\i}sica, casilla  487-3 Santiago, Chile. } }

\address{ Universit\'e de Gen\`eve,  D\'epartement de Physique Th\'eorique,
          24, Quai Ernest-Ansermet, CH-1211 Gen\`eve 4, Switzerland}
\date{2nd December, 1996}
\maketitle
\begin{abstract}
 Continuous One-dimensional models  supporting extended states
  are studied. These delocalized
 states
 occur at well defined values of the energy and  are consequences
 of simple statistical correlation rules.    We  explicitly study alloys
   of $\delta$-barrier potentials as well as  alloys and liquids of quantum wells.
 The divergence of the localization length is studied and a critical
 exponent $2/3$  is found for the $\delta$-barrier case, whereas for the quantum wells 
 we find an exponent of $2$ or $2/3$ depending on the well's parameters. 
  These results support the idea that
 correlations  between random scattering sequences  break
 Anderson
 localization. We further calculate the conductance of disordered 
superlattices. At the peak transmission the relative fluctuations of 
the transmission coefficient are vanishing.
\end{abstract}

\begin{section}*{  Introduction}

Recently the interest has increased substantially in order to find theoretical evidence for the 
breaking of Anderson localization by internal correlations in disordered systems
\cite{flores,dunlap1,dunlap2,dunlap3,bovier,gango,flores2,datta,dominguez1,diez,dominguez2,dominguez3,evangelous,evangelous2,evangelous3,hilke,heindrich}. For instance, in ref.\cite{flores} a tight-binding Hamiltonian
with site correlations between the diagonal and the off-diagonal potentials
was studied.  A divergence of the localization length was obtained 
with a critical exponent $\nu=2$ outside the band edges.
In Refs. \cite{dunlap1,dunlap2,dunlap3,bovier,gango} a simple discrete
 model with
correlations expressed by pairing (dimer) was studied. Their main result was the 
existence of a divergent localization length at some critical energies. 
In fact,  this
divergence   is related to the existence of  delocalized states
found explicitly in ref. \cite{flores2}, which can be related to random phases and thus to
extended states. Similar dimer models for classical systems were studied
in ref. \cite{datta}. Using the inter-relation between  a
disordered Kronnig-Penney system  and the  dimer  model (Poincar\'e map)
\cite{dominguez1,diez,dominguez2,dominguez3}   an infinite set
of such delocalized states exist. The generalization of the dimer
model to the N-mer case
  was considered, for instance, in
refs. \cite{evangelous,evangelous2,evangelous3,hilke}. In this context  perturbative
methods were used in ref. \cite{heindrich}.

 In this way,  delocalized states have been found
`contradicting'  the usual belief that all eigenstates are localized in
one-dimension (1-d).
In fact, this is only  apparent because the theorems related to localization
in 1-d consider strictly uncorrelated random potentials (see for instance
Ref.\cite{mott,ishii,delyon,kunz,johnston}). Interesting numerical simulations
were carried out for binary
disordered systems in ref.\cite{davids}.

Usually these correlated disordered systems support delocalized states for
well defined energies. In this respect,  a `band of delocalization'  can be
defined when the length $L$  of the sample is smaller than  the
 localization length  $L_c(E)$, which diverges near  the critical energy $E_c$.
    Quantitatively,  the region of delocalization can be
  defined approximately, as
  \begin{equation}
  L<L_c(E)\sim \frac{1}{|E-E_c|^\nu},
  \end{equation}
  where $\nu$ has been  determined explicitly in some models.
Evidently this definition of a `band of delocalization'  is controversial
because it depends on the existence of states in the interval $
|E-E_c|$ and, moreover, is very different of the usual one related
to true delocalized (unnormalizable) states.  For instance,
in the dimer model \cite{dunlap1,dunlap2,dunlap3} a band of delocalized states 
is found  to follow  as $\Delta E\sim \sqrt L$, where $L$ is proportional to 
 the number 
of impurities $N$,   nevertheless, the total number of states
grows as $L$ and therefore the relative number of delocalized states tends to zero
as $1/\sqrt L$. It has been furthermore recently  shown that additional 
perturbations on the dimer potentials lead to a suppression of these 
extended states \cite{evan3}. Therefore these delocalization properties 
become more relevant in finite systems, as presented in the last part of this 
article. In quasi-one-dimensional systems, like  models with dimerized 
interchain couplings \cite{evan4}, leading   
to a cantor-set-like conducting band, the finite temperature transmission  can 
be greatly enhanced by the resonance energies.

As noticed, internal correlations break strong localization. In this
paper  we consider some simple models showing delocalization properties.
For a general point of view we consider the one-dimensional
 disordered Hamiltonian of a particle in a random potential
\begin{equation}
H=\frac{p^2}{2m}  +\sum_l  V_l(x-x_l),
\end{equation}

\begin{equation}
V_l(x)= \left\{  
\begin{array}{cc} \neq 0 & -a_l<x<a_l \\
  =0 & \mbox{otherwise}   \end{array}  \right. ,
\end{equation}
thus, a sequence of scattering barriers centered at the random position
 $x_l$ without overlap. Every barrier is symmetric with random
 support $2a_l$ and the distance between two barriers is   given by
 \begin{equation}
 x_{l+1} -x_l=d_l,
 \end{equation}
 which is a  random quantity.

 Between consecutive barriers the particle propagates freely and
 its  wavefunction  can be expressed as
 \begin{equation}
 \psi_l(x)= A_le^{ikx} +B_le^{-ikx}\quad \mbox{for}\quad  x_l+a_l<x<x_{l+1} -a_{l+1},
 \end{equation}
where $k$ is the wavenumber related to the energy by $E=\hbar^2k^2/2m$.
Theoretical group arguments \cite{merbacher,erdos,felderhof} relate  the amplitudes $(A_{l+1},
B_{l+1})$  to $(A_l,B_l)$ by means of
\begin{equation}
\left( \begin{array}{c}  A_{l+1}\\B_{l+1} \end{array} \right) =
\left( \begin{array}{cc} e^{-ikx_l}&0\\0&e^{ikx_l} \end{array} \right)
\left( \begin{array}{cc} \alpha_l&\beta_l\\ \beta^*_l &\alpha^*_l \end{array}
        \right)
\left( \begin{array}{cc}  e^{ikx_l} &0 \\0&e^{-ikx_l}  \end{array}  \right)
\left( \begin{array}{c} A_l\\B_l \end{array} \right)
\end{equation}
where the elements $\alpha,\beta$ of the collision matrix  are inter-related
like $|\alpha |^2-|\beta |^2=1$  and we further on assume
$\beta^*=-\beta$ (symmetric barrier).
This form of the collision matrix is related to the fact that the
time-independent  Schr\"odinger equation is real.

Defining the wavefunction  just after the collision with the  $l-$th barrier as
\begin{equation}
\psi_l^+ =A_le^{ik(x_l+a_l)}  +B_le^{-ik(x_l+a_l)},
\end{equation}
a condition of delocalization is 
\begin{equation}
\psi_{l+1}^+=\pm \psi_l^+,
\end{equation}
which relates the elements of the collision matrix as
\begin{equation}
\alpha_le^{ik(d_l+a_{l+1}-a_l)} +\beta_l^*e^{-ik(d_l+a_{l+1}+a_l)}
=\pm 1.
\end{equation}
The above delocalization condition is a sufficient one and, as expected,
 is  not verified for  totally uncorrelated systems. Condition (9) yields using 
 $|\alpha |^2-|\beta |^2=1$ the 
 following  
  simple delocalization condition 
 \begin{equation}
 |\alpha_l|=1 \quad \mbox{ and }\quad \beta_l=0,
 \end{equation}
  which leads to $|A_{l+1}|=|A_l|$ and $|B_{l+1}|=|B_l|$. 

Condition (9)  inter-relates the random parameters like $a_l$ and $d_l$  in the disordered
system and  is therefore a source of correlation between  these
random parameters.
In fact, in some simple systems,  relations (9) and (10) can be  satisfied and we present two
examples: The $\delta$-barriers-alloy
 and the quantum well liquid-alloy with interrelated random parameters.

 To end this section, we remark that any relation similar to (6) can be
 formally written   as a tight binding Schr\"odinger  equation
 (Poincar\'e map) by considering
 \begin{equation}
 \hat \alpha_l =\alpha_le^{ik(d_l+a_{l+1}-a_l)},\quad
 \hat \beta_l  =\beta_le^{ik(d_l+a_{l+1}+a_l)}
 \end{equation}
 and 
 \begin{equation}
 D_{l+1} \psi_{l+1}^++D_l\psi_{l-1}^+  =
 V_l \psi_l^+,
 \end{equation}
 where
 \begin{equation}
 D_l=\frac{D(k)}{ 2i(Im \hat \alpha_l  -Im \hat \beta_l )}  \quad \mbox{ and } \quad 
 V_l=\{ D_{l+1}(\hat \alpha_l+\hat \beta_l^*) +D_l (\hat \alpha_{l-1}^*
 -\hat \beta_{l-1}^*) \}. 
 \end{equation}
 We notice that the above expression is only a formal one. For models such
 as the $\delta$-barriers sequence, $D_l$ is well defined
 for any $k$ with the arbitrary   
 choice $D(k) =2i\sin k$. Nevertheless, in  other systems
  singularities for  some values of the wavenumber $k$ can appear.

 \end{section}

 \begin{section}*{ I. Delocalization in a $\delta$-barrier sequence}

 Consider a sequence of $\delta$-potentials which are statistically
 distributed over  lattice sites  (alloy). Thus consider the sequence
 \begin{equation}
 V_l(x)=V_l\delta (x-x_l)\quad \mbox{and} \quad  d_l=d,
 \end{equation}
 where $V_l$ are random uncorrelated parameters and $d$ is a  constant
 lattice parameter. In this case, the elements of the collision matrix are given
 by
 \begin{equation}
 \alpha_l=1+i\frac{V_l}{2k}\quad \mbox{and} \quad   \beta_l=i\frac{V_l}{2k}.
 \end{equation}
 The delocalization condition (9) becomes explicitly
 \begin{equation}
 e^{ikd}  -\frac{V_l}{k}\sin (kd)  =\pm 1   .
 \end{equation}
In general, for any arbitrary momentum $k$, this condition does not hold
because of the random quantity $V_l$. Nevertheless,  if $k=n\pi/d$
( $n\in Z^*$), as first observed by Ishii \cite{ishii},
  we have a set of delocalized states where
$|\psi_{l+1}|=|\psi_l|$.  This is at first sight very surprising  because of the simplicity
of the disordered model. Evidently, correlations exist  and 
are related to the choice $d_l=d$ for any $l$. In fact, assuming a sample
with $N$ barriers, the usual
$2N$  random parameters characterizing the  uncorrelated
system are reduced to $N$ because of the constraint $d_l=d$.

In this model the physical interpretation is simple, in fact as long as $k=n\pi/d$
the electron does not `feel' the random potential because the distance
$d$, between consecutive barriers, is a multiple of its wavelength.

At this point, it is interesting to study the divergence of the
localization length $L_c(E)$ near to the critical energy  $E_c=\frac{1}{2m}
( \frac{\hbar n\pi}{d} )^2 $. Using the Poincar\'e map (12) for this
model we have
\begin{equation}
\psi_{l+1}+\psi_{l-1} =2\{  \cos kd  \frac{V_l}{k} \sin kd \} \psi_l.
\end{equation}
For $\epsilon=E_c-E\ll1$ and $\hbar/2m$ taken as unity,
 equation (17) can be rewritten 

\begin{equation}
\psi_{l+1}+\psi_{l-1}=\pm\left(2- V_l\epsilon\frac{d^3}{n^2\pi^2}\right)\psi_l.
\end{equation}
 This last model was extensively studied in the 
limit $\epsilon\ll1$ by 
Derrida and Gardner \cite{derrida}. They calculated the 
complex Lyapounov exponent $\gamma$, where the real part corresponds to the 
inverse localization 
length and the imaginary part to $\pi$ times the integrated density of 
states. Their results 
can be expressed as follows:
\begin{equation}
\left\{\begin{array}{l}
Re(\gamma) \simeq K_1\epsilon^{2/3}\langle V_n^2\rangle ^{1/3}\frac{d}{(n\pi)^{2/3}}\\
Im(\gamma) \simeq K_2\epsilon^{2/3}\langle V_n^2\rangle ^{1/3}\frac{d}{(n\pi)^{2/3}},
\end{array}\right. 
\end{equation} 
where $K_1=0.29\dots $ and $K_2=0.16\dots $ and $\langle\cdot\rangle$ is 
the average over all 
impurities.
From (19) it is straightforward that the inverse localization length $L_c^{-1}$ scales as 
\begin{equation}
L_c^{-1}\sim\epsilon^{2/3}\langle V^2\rangle ^{1/3}
\end{equation} 
and the density of states is 
\begin{equation}
\rho(\epsilon)=\partial_{\epsilon}Im\gamma(\epsilon)\sim\epsilon^{-1/3}.
\end{equation}

The main result of this section is to show that a continuous disordered 
model where the potential amplitudes are random but 
 placed on a lattice, exhibits extended states for 
discrete values of the energy. We recall that for this model the result 
was first pointed out by Ishii \cite{ishii}.
However, the exponent $2/3$ we obtain is different to the $1/2$
obtained by Ishii, who considered a Cauchy distribution. 
The correlation in this model comes from the 
fact that the impurities are placed on a regular lattice. This in fact is 
a very natural physical assumption, as one can imagine an alloy where 
different atoms are placed randomly, described by   the random amplitude of 
the potential, but on a lattice. This system would lead to a similar 
model studied above. In fact the models where the locations of the impurities 
are continuously random, no extended states have been found so far for delta like potentials.
For finite potentials they do exist as we will mention in the next section.

\end{section}
\begin{section}*{ II. Delocalization in a quantum well sequence}

The case of delta impurities considered in the previous section is very special and 
cannot be used in order to describe more general impurity potentials. This is why it is 
of interest to study other types of potentials, which could eventually describe more general 
impurity potentials. One of the simplest examples of such a potential is the rectangular well 
potential. In principle one could approximate any potential with rectangular wells if the 
elementary width is small enough. 
Therefore we will restrict our considerations to
rectangular well potentials.

As an important result we observe the existence of extended states for systems where we have 
continuous spatial disorder and discrete shape disorder. This has to be contrasted with the 
case of delta impurities where extended states exist only when we have shape disorder alone, i.e., 
where the amplitudes of the impurity potentials are random.

The aspect and definitions of the rectangular well potential are shown in fig. 1:


\begin{figure}[h]
\begin{minipage}{16cm}
\unitlength1cm
\hspace*{3cm}
\begin{picture}(10,2)
\thicklines
\put(0,0){\line(1,0){2}}
\put(2,0){\line(0,1){1}}
\put(2,1){\line(1,0){3}}
\put(5,1){\line(0,-1){1}}
\put(5,0){\line(1,0){1}}
\put(6,0){\line(0,-1){2}}
\put(6,-2){\line(1,0){2}}
\put(8,-2){\line(0,1){2}}
\put(8,0){\line(1,0){2}}
\put(3.5,0){\circle*{0.1}}
\put(7,0){\circle*{0.1}}
\put(3,-1){\makebox(1,1){$x_l$}}
\put(6.5,-1){\makebox(1,1){$x_{l+1}$}}
\thinlines
\put(4.5,.5){\vector(1,0){.5}}
\put(4,.5){\vector(-1,0){.5}}
\put(4,.25){\makebox(.5,.5){$a_l$}}
\put(5.25,-1.25){\makebox(.5,.5){$d_l$}}
\put(5.25,-1){\vector(-1,0){1.75}}
\put(5.75,-1){\vector(1,0){1.25}}
\put(1.5,0.5){\vector(0,-1){.5}}
\put(1.5,.5){\vector(0,1){.5}}
\put(1,.25){\makebox(.5,.5){$V_l$}}
\end{picture}
\vskip 2.5cm
\caption{Rectangular well potentials}
\end{minipage}
\end{figure}
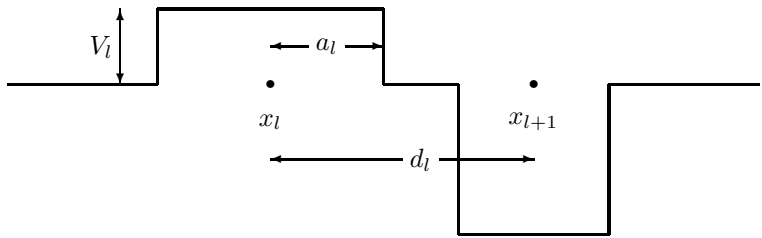
\vskip .5cm

For this case we have

\begin{equation}
\alpha_l=\frac{e^{-i2a_lk}}{2p_lk}\left(2p_lk\cos(2a_lp_l)+i(p_l^2+k^2)\sin(2a_lp_l)\right)
\end{equation}
and
\begin{equation}
\beta_l=i\frac{(p_l^2-k^2)\sin(2a_lp_l)}{2p_lk},
\end{equation}
where $p_l=\sqrt{|k^2-V_l|}$.
 We have to point out that these results are 
only valid when $E\geq V_l$. For the case $E\leq V_l$ we have to replace $cos(\cdot)$ and 
$sin(\cdot)$ in (22,23) by their hyperbolic counterparts. 

Using the delocalization condition (10) we have an extended state when  $\beta_l=0$, thus
\begin{equation}
E=k^2=V_l+\frac{n_l^2\pi^2}{4a^2_l}\quad\mbox{with $n_l$ an integer}.
\end{equation}
This means that if we have a discrete set of impurity configurations, we have critical energies 
$E_c$ at which the states are delocalized. The impurities have just to be constructed with 
the help of equation (24) once we replace $E$ by $E_c$, where we can take different values 
for $V_l$ or $a_l$ by changing $n_l$. A {\em very remarkable} aspect of these disordered 
models is that they are {\em independent} of the space disorder expressed by $d_l$ as was 
already appreciated by Tong \cite{tong} for a different model. 

We evaluate the exponent for the following case: suppose $d_l=d$ and $a_l=a$, i.e., we are 
left with a pure shape disorder. Let $k_c^2=E_c$ be the critical energy defining the correlation 
(24) of the disorder and writing the Poincar\'e map (12) around $k=k_c+\epsilon$, with $\epsilon$ 
very small, we obtain 
\begin{equation}
\psi_{l+1}+\psi_{l-1} =\{  2\cos \left(k_c(d-2a)\right) +\epsilon\xi_l \} \psi_l,
\end{equation}
where $\xi_l$ is random and function of $V_l$. For the case where $k_c(d-2a)/\pi$ is an integer 
we obtain the same result as for the delta-barrier case (18-21), thus $\nu=2/3$. When 
$k_c(d-2a)/\pi$ is not an integer we can also use Derrida and Gardner's \cite{derrida} result for 
the case inside the band, yielding $\nu=2$. The most general case, where $d_l$ and $a_l$ are 
random, and when no particular configuration is used, we obtain numerically the exponent $\nu=2$. 

The interesting aspect of these 
continuous disordered systems is that they can describe a real physical situation of electrons in a certain 
potential.  The problem with continuous models is that the expressions of the transfer 
matrices can be much more complicated except if one considers delta impurities or rectangular well 
potentials. For more general impurity potentials one can either approximate them with a sum 
of rectangular barriers or use the semiclassical WKB approximation. This last approximation can, however, 
have dramatic effects on the localization properties, especially around some
 critical energies,  
as they are very sensitive to small perturbations.  

Summerizing the localization length divergences for the $\delta$-impurity and 
quantum wells we have:

  \begin{equation}
  L_c(E)\sim \frac{1}{|E-E_c|^\nu} \mbox{ , where }\nu=\left\{  
\begin{array}{ll} =2/3 & \mbox{$\delta$-barrier and $k_c(d-2a)/\pi$ 
 an integer for the quantum well} \\
  =2 & \mbox{otherwise}   \end{array}  \right.
  \end{equation}

\end{section}
\begin{section}*{III. Fluctuations and conductance}
In one dimensional disordered systems, the relative fluctuations of the 
transmission coefficients diverge with the size of the system 
\cite{felderhof}. Therefore 
the average transmission is a bad statistical quantity. However at the 
critical energies discussed above, the states are deterministic and 
therefore one would expect no fluctuations. In the following we 
evaluate the relative fluctuations explicitly and calculate the conductance 
for superlattices around these critical energies. 

Around the critical energy $E_c$  we evaluated the localization length
 dependence on energies, therefore 
 $T_N\sim e^{-L/L_c}$, where $T_N$ is the transmission coefficient after
 $N$ barriers and $N$ is proportional to $L$. The relative fluctuations are then given by
\begin{equation}
\frac{\Delta T}{T}\sim N\frac{\Delta L_c}{L_c^2}.
\end{equation}
Using the central limit theorem, valid for the logarithm of the 
product of random determinant one matrices, 
the relative 
fluctuations of the localization length can be expressed as 
\begin{equation}
\frac{\Delta L_c}{L_c}\sim N^{-1/2},
\end{equation} 
hence using (26-28) 
\begin{equation}
\frac{\Delta T}{T}\sim N^{1/2}|E-E_c|^\nu.
\end{equation}
In usual one-dimensional disordered systems the average distance between 
eigenvalues is 
of the order $N^{-1}$, therefore using $|E-E_c|\sim N^{-1}$, 
 we obtain 
\begin{equation}
\frac{\Delta T}{T}\sim |E-E_c|^{\nu-1/2}.
\end{equation}
This demonstrates that for the case of interest here, both 
cases $\nu=2$ and 
$\nu=2/3$ have vanishing relative fluctuations near the critical 
energies. This fluctuation analysis gives also a bound on the 
possible divergence exponent $\nu$. In fact as long as $\nu>1/2$, 
the average transmission around the critical energies is a good 
statistical quantity.

Many 
 physical systems  can be very well approximated with either 
rectangular well potentials or delta potentials. This is the case for 
superlattices of 
heterostructures. One can grow very precise, up to the atomic precision, 
layers of $GaAs$. 
Then one dopes differently one layer with $Al$ and then again the same layer 
and one 
obtains in this way a single so called delta doped layer. This can then be 
repeated as often 
as one wishes. This last structure is very well described in the direction of 
the layers 
by delta impurity models.  Instead of doping differently only one layer we 
can also dope a finite 
number 
of layers differently. In this case the potential looks more like a 
rectangular well potential, 
where the amplitude depends on the concentration of the doping used.  
In this way it is possible to grow one-dimensional quantum wires where both 
directions of the layers are etched, two-dimensional systems where only one 
of the direction of the 
layers is etched and finally three-dimensional systems. 

The two probe  conductance is simply given 
by the sum of the transmission coefficients \cite{fisher}, thus
\begin{equation}
G(\mu)=\sum_{k|E=\mu}T(k)\cdot \frac{e^2}{h},
\end{equation}
where $\mu$ is the chemical potential.

For a one-dimensional quantum wire the conductance is just given by the 
transmission coefficient, as there is only one conducting channel. Around the critical 
energies   we evaluated the localization length dependence on energies, therefore 
using $T_N\sim e^{-N/L_c}$, where $T_N$ is the transmission coefficient after $N$ barriers,
we obtain  conductance peaks at the 
critical energies of the 
form 
\begin{equation}
G(\mu)\sim e^{-\gamma|\mu-E_n|^{\nu}},
\end{equation}
where $\mu$ is the chemical potential, $\gamma$ depends on the size and 
disorder of the system and 
$E_n$ are the critical energies. For the delta doping $E_n=\frac{1}{2m}
( \frac{\hbar n\pi}{d} )^2 $ and $\nu=2/3$. For the rectangular case 
we have one critical energy $E_c$ and the random $Al$ doping potentials are 
given by 
\begin{equation}
V_l=E_c-\frac{n_l^2\pi^2}{a^2}\quad\mbox{with $n_l$ a random integer}
\end{equation}
and $a$ is the width of the doped layers. The width $d_l$ of the undoped 
layers can 
be random and 
we obtain $\nu=2$. For the special case where $d_l=d$ is constant the value 
of the exponent $\nu$ depends on the ratio $E_c(d-2a)^2/\pi^2$. If this ratio
is an integer the exponent is $\nu=2/3$ but remains $\nu=2$ otherwise.
The interesting feature of this system, is that it would behave as 
a perfect filter. The band-width could be controlled by the size or $\gamma$ 
and the desired critical energy by the choice of dopants.

In order to evaluate the conductance of the three-dimensional system, as 
defined above, we 
have to sum over all possible channels. In the directions (x,y) perpendicular 
to the grown layers the density of states is simply the usual box filling 
density. In the infinite limit we have to integrate over the Brillouin zone
which yields,
\begin{equation}
G\sim\int_{k_x^2+k_y^2\leq\mu}e^{-\gamma|\mu-k_x^2-k_y^2-E_c|^\nu}dk_xdk_y
\end{equation}
 The plots of the conductance from eq. (34) 
as a function of the chemical potential 
are given in fig. 2 for different system sizes characterized by $\gamma$ 
and for 
the two cases $\nu=2$ and $\nu=2/3$.

The two-dimensional case is presented in fig. 3. In this case the behavior 
is consequently different for  $\nu=2$ and $\nu=2/3$. Indeed for 
$\nu=2/3$ we have a singular derivative, which could be observed 
experimentally, in order to distinguish between the two cases.

\input epsf
\begin{figure}
\hskip 1.5cm
\epsfxsize=6cm
\hspace*{0cm}
\epsfbox{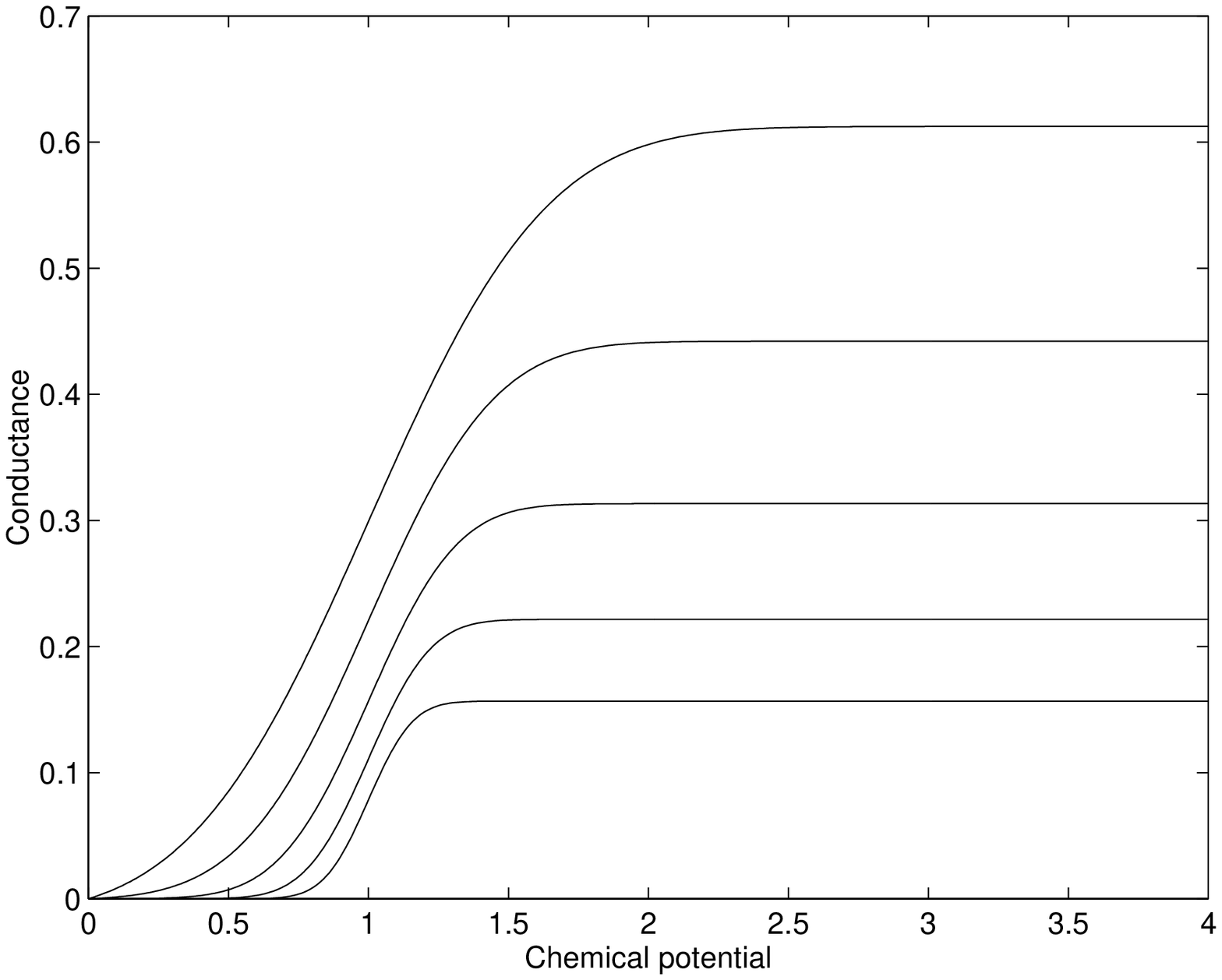}
\hspace*{1.5cm}\epsfxsize=6cm\epsfbox{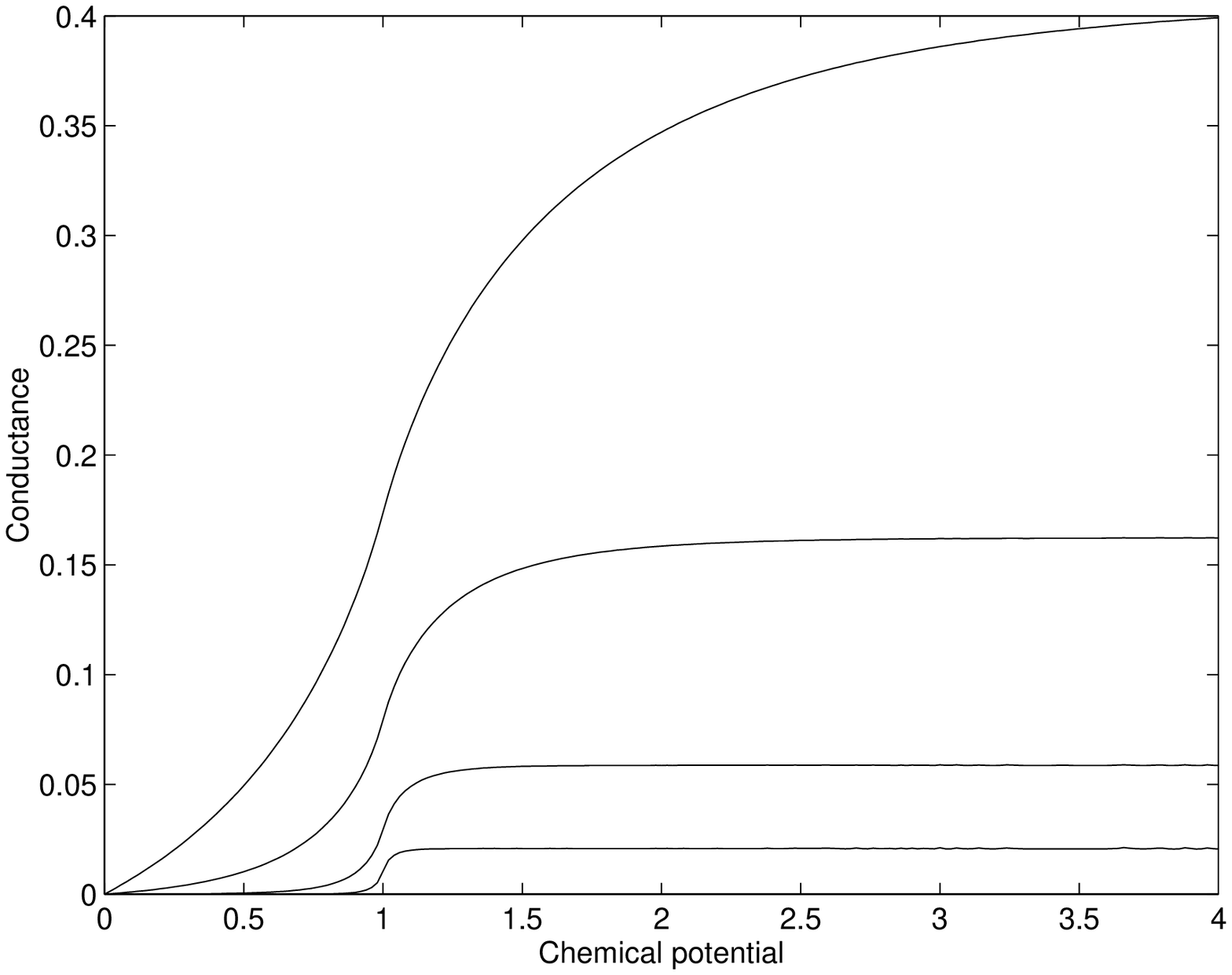}
\vskip .3cm
\caption{Conductance for $\nu=2$ and $\nu=2/3$ (right graph) 
as a function of the chemical potential, for the
case where the resonance energy is $E_c=1$. The different curves represent
different values of $\gamma=2^n$, with $n=1,2,3,4,5$. The uppermost
curve represents the case $\gamma=2$. }
\end{figure}

\input epsf
\begin{figure}
\hskip 1.5cm
\epsfxsize=6cm
\hspace*{0cm}
\epsfbox{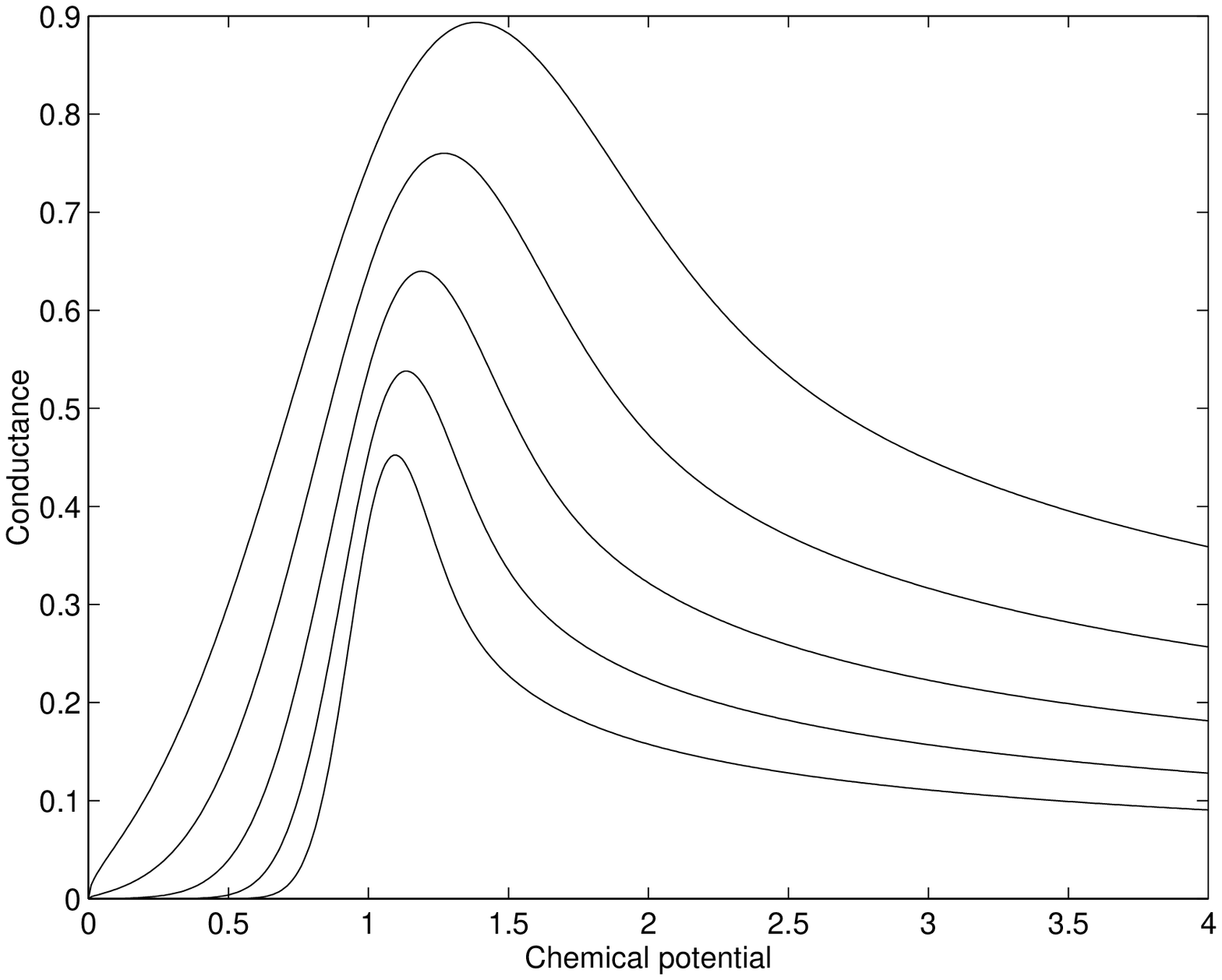}
\hspace*{1.5cm}\epsfxsize=6cm\epsfbox{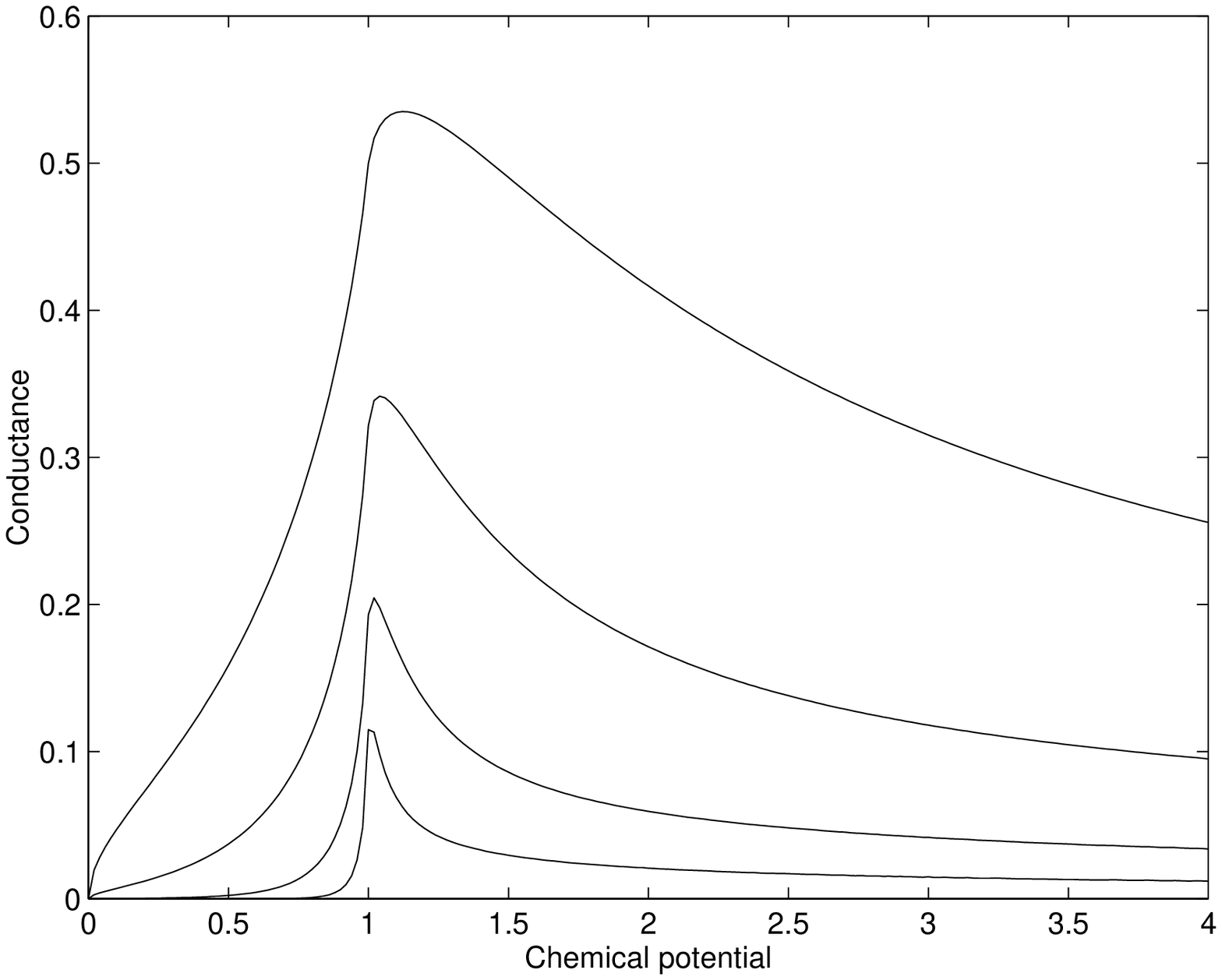}
\vskip .3cm
\caption{Conductance for the two-dimensional case with the same parameters 
as in fig. 2.}
\end{figure}

The quantitative values 
can also be estimated. In the delta doping case the typical Fermi wavelength 
is of the 
order of 200{\AA} and the typical layer thickness is of the order of 5{\AA}. 
This means that 
we have to construct delta doped layers with around $40$ layers in between. 
In order to obtain 
the disorder we just have to chose randomly the dopant or the concentration 
of the doping for 
the delta layers.  The same reasoning can be held for rectangular 
potentials, where one just has to chose a particular potential configuration.

\end{section}

\begin{section}*{ Conclusions}
   In the first section we demonstrated the existence of extended states in continuous disordered 
systems with delta impurities placed on a lattice and obtained for the exponent describing  
the divergence of the localization length $\nu=2/3$. This model hence demonstrates the 
existence of extended states in systems with only shape disorder. In the second section,  
using rectangular potentials we demonstrated that models with shape {\em and} space disorder can 
also exhibit extended states. The only constraint on the disorder distribution is that it must 
be discrete for the shape disorder but can be continuous for the space disorder. 
In this case the correlation in disorder leading   
to the existence of delocalized states is simply the fact that the distribution mentioned 
above is discrete and obeys relation (24). 
Therefore one has to be very careful when one discusses disordered systems 
using discrete disorder distributions as there can exist singularities in the spectrum. 
The exponents of the localization length divergence are very interesting. In fact, depending on 
the parameters of the quantum wells we obtain different exponents, either $\nu=2/3$ or $\nu=2$.

The relative fluctuations of the conductance around these critical 
energies are vanishing. Therefore the conductance is a 
well defined statistical average and represents the transport properties 
of these disordered systems.

\end{section}

 \subsection*{Acknowledgements}

 J. C. F. is supported by grant FONDECYT-CHILE  3940004 and M. H. by the Swiss National Science 
 Foundation.


\end{document}